\documentclass[11 pt,final,journal,letterpaper,oneside,onecolumn]{IEEEtran}% The IEEEtran.cls file specified here.
\usepackage{graphicx}%               % need for figures
\usepackage{amssymb,amsmath,amsfonts}% need for subequations, \align
\usepackage{suffix}
\usepackage{mathtools}
\usepackage{cite}  
\DeclarePairedDelimiterX\MeijerM[3]{\lparen}{\rparen}%
{\begin{smallmatrix}#1 \\ #2\end{smallmatrix}\delimsize\vert\,#3}

\newcommand\MeijerG[8][]{%
  G^{\,#2,#3}_{#4,#5}\MeijerM[#1]{#6}{#7}{#8}}

\WithSuffix\newcommand\MeijerG*[7]{G^{\,#1,#2}_{#3,#4}\MeijerM*{#5}{#6}{#7}}                  % use for citing multiple references within the same brackets
\ifCLASSINFOpdf
\else
\fi
\begin{document}
%
% paper title
% Titles are generally capitalized except for words such as a, an, and, as,
% at, but, by, for, in, nor, of, on, or, the, to and up, which are usually
% not capitalized unless they are the first or last word of the title.
% Linebreaks \\ can be used within to get better formatting as desired.
% Do not put math or special symbols in the title.
\title{Secrecy Outage for Wireless Sensor Networks}

% author names and IEEE memberships
% note positions of commas and nonbreaking spaces ( ~ ) LaTeX will not break
% a structure at a ~ so this keeps an author's name from being broken across
% two lines.
% use \thanks{} to gain access to the first footnote area
% a separate \thanks must be used for each paragraph as LaTeX2e's \thanks
% was not built to handle multiple paragraphs
%

%\author{Furqan Jameel, Shurjeel Wyne,~\IEEEmembership{Senior Member,~IEEE}}%
\author{Furqan Jameel, Shurjeel Wyne, and Ioannis Krikidis}%
\maketitle

% As a general rule, do not put math, special symbols or citations
% in the abstract or keywords.
\begin{abstract}
This paper addresses Physical Layer Security (PLS) for a wireless sensor network in which multiple sensor nodes communicate with a single sink in the presence of an eavesdropper. Under the assumption of sink using outdated channel state information (CSI) of the eavesdropper link, the outage probability of achievable secrecy rate under Weibull fading is investigated and its analytical expressions are derived. Specifically, we consider sensor scheduling to enhance PLS, thereby we quantify the degradation in outage probability due to outdated CSI under the best-node scheduling scheme and compare it with conventional round-robin scheduling. Finally, we analyze secrecy outage probability conditional on successful transmission and provide practical insights into the effect of outdated CSI of the eavesdropper. The derived theoretical results are validated through simulations.
\end{abstract}

% Note that keywords are not normally used for peerreview papers.

\IEEEpeerreviewmaketitle

\section{Introduction}
Wireless sensor networks (WSNs) are widely used for military applications such as navigation and surveillance \cite{zou2015improving} as well as for industrial automation \cite{zou2016intercept}. Link security is a critical aspect of successful WSN operation. Traditional cryptographic techniques are not suitable for securing WSNs, because they require hardware complexity and consume large amounts of energy that are not affordable in a WSN \cite{zou2015improving}. Moreover, an eavesdropper with unlimited computing power may still decipher these techniques using brute-force attack. In this context, Physical Layer Security (PLS) has emerged as an attractive solution for securing wireless transmissions by exploiting the wireless channel characteristics \cite{qin2011optimal}. Since PLS techniques such as artificial noise generation do not suit WSNs due to their limited energy resources \cite{qin2011optimal}, sensor scheduling has been proposed as a less energy-intensive scheme for WSN security \cite{zou2016intercept}.

In recent years, the Weibull distribution has been successfully used to model small-scale fading (SSF) measured in indoor and outdoor scenarios of practical significance, see for example \cite{wu20105,ibdah2015mobile} and references therein. In \cite{wu20105}, the authors refer to Vehicle to Vehicle channel measurements in the 5 and 10-GHz band; they propose to model the SSF with the Weibull distribution with its shape parameter $\beta$ between 1.77 and 3.9 for the 5-GHz band, and in the range 2.02-3.95 for the 10-GHz band. In \cite{ibdah2015mobile}, the authors reported on mobile-to-mobile channel measurements at 1.85 GHz in a dense-scattering suburban scenario. They observed that the SSF is accurately modeled by the Weibull with its $\beta$ taking values between 1.16 and 1.5; note that decreasing $\beta$ values model increasing fading severity. Despite this practical significance of the Weibull fading model, PLS under Weibull fading has not been sufficiently investigated in the literature. Recently in \cite{liu2016average}, an expression for the average value of the secrecy capacity (SC) under Weibull fading was derived but no practical scheme was proposed to improve the secrecy performance. Moreover, most PLS investigations consider the ideal case that perfect channel state information (CSI) is available for the eavesdropper channel\cite{zou2016intercept, liu2016average}. The submitted work addresses these limitations by providing a complete characterization of the secrecy outage performance with outdated eavesdropper CSI. Specifically, the main contributions of this work are:
\begin{itemize}
	\item Derivation of an analytical expression for the secrecy outage probability under Weibull fading when outdated channel state of eavesdropper link is used by the transmission scheme; 
	\item Analysis of the outage performance under Weibull-fading of best-node scheduling with imperfect eavesdropper CSI and round-robin scheduling, whereas \cite{zou2016intercept} treats the Nakagami-$m$ fading case with perfect CSI of all links;
	\item Investigation of secrecy outage probability conditional on successful decoding and evaluation of the impact of outdated eavesdropper CSI when used by the transmission scheme.
\end{itemize}
The remainder of this paper is organized as follows. The system model is given in Sec. II; Sec. III contains derivation and analysis of the outage expressions. In Sec. IV numerical results are provided. Finally, Sec. V concludes this work.

\section{SYSTEM MODEL}

\begin{figure}
\centering
	\includegraphics[scale=0.25]{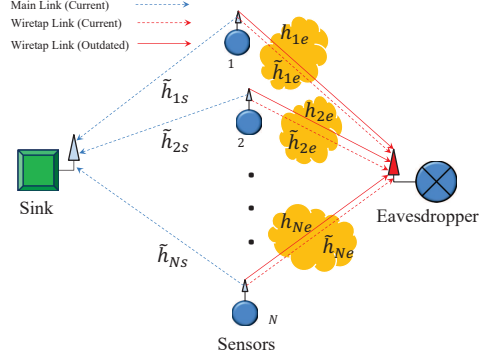}
	\caption{System Model.}
	\label{fig:figblock}
\end{figure}

We consider a WSN consisting of $N$ sensor nodes transmitting over orthogonal channels to a sink in the presence of an eavesdropper as shown in Fig \ref{fig:figblock}. The main-links experience statistically independent identically distributed flat Weibull fading, whereas the Weibull fading on wire-tap links, between sensors and the eavesdropper, is not necessarily identically distributed. The sink as the supervisory node facilitates communication among different nodes through feedback; hence it is assumed to have CSI for the $N$ sensor channels as well as the eavesdropper channels \cite{zou2013optimal}. To be more specific, each sensor estimates its own CSI and transmits this information to the sink \cite{zou2016intercept} under the assumption of reciprocal channels. The eavesdropper itself is assumed to be a legitimate receiver in the network for some signals and acts as an eavesdropper for others as in a multicast and unicast scenario, respectively \cite{wang2014secure,huang2015secure}. Under these conditions the eavesdropper's CSI estimated during its legitimate reception may become outdated during its eavesdropping activity. This condition is incorporated into our analysis by introducing a correlation model with coefficient $0\leq\rho_e\leq1$ between the currently valid CSI of the eavesdropper, which is unavailable to the sink, and its previously estimated value that has become outdated but is available to the sink; perfect CSI is assumed elsewhere. After collecting the CSI from all sensors, the sink can select a particular node for transmission and subsequently notify the network. Let $S=\left\{S_i|i=1,2,...N\right\}$ denote the set of $N$ sensor nodes. Consider that $S_i$ transmits its signal $x_i$ to sink with power $P_i$. Then the received signal at sink is written as
%%%%%%%%%%%%%%%%%%
%%%%%%%%%%%%%%%%%% 
\begin{equation}
y_{is}=\sqrt{P_i}\tilde{h}_{is}x_i+n_s,
\end{equation}
%%%%%%%%%%%%%%%%%%
where $\tilde{h}_{is}$ represents the channel between $S_i$ and sink with $|\tilde{h}_{is}|$  Weibull-distributed. Furthermore, $n_s$ represents the zero mean additive white Gaussian noise (AWGN) at sink with variance $N_0$. The instantaneous signal-to-noise ratio (SNR) of the $i$-th main link is written as $\tilde{\gamma}_{is}=\frac{|\tilde{h}_{is}|^2P_i}{N_0}$. Since the transmission from $S_i$ is also picked up by the eavesdropper, the signal received at eavesdropper is given as
%%%%%%%%%%%%%%%%%%
\begin{equation}
y_{ie}=\sqrt{P_i}\tilde{h}_{ie}x_i+n_e,
\end{equation}
%%%%%%%%%%%%%%%%%%
where $n_e$ is the AWGN at eavesdropper with variance $N_0$, and $\tilde{h}_{ie}$ represents the channel between $S_i$ and the eavesdropper with $|\tilde{h}_{ie}|$ Weibull-distributed and instantaneous SNR of $\tilde{\gamma}_{ie}=\frac{|\tilde{h}_{ie}|^2P_i}{N_0}$. However, this CSI is not available to the sink that only has information of the outdated CSI of the wiretap link ${h}_{ie}$, which is related to the current but unknown CSI $\tilde{{h}}_{ie}$ through a correlation model  ${h}_{ie}=\rho_e\tilde{h}_{ie}+\sqrt{1-\rho_e^2}v$ \cite{hu2016versatile}; here $v=\mathcal{CN}(0,1)$ is channel independent noise. The outdated SNR of the wiretap link known at the sink is written as ${\gamma}_{ie} = \frac{|{h}_{ie}|^2P_i}{N_0}$. Under these conditions, the joint distribution of the current and outdated wiretap SNRs is given by the joint distribution of two correlated Weibull random variables $V_1$ and $V_2$ written as \cite{sagias2005correlated} 
%%%%%%%%%%%%%%%%%%
\begin{align}
f_{V_{1},V_{2}}(y_{1},y_{2})&=\beta ^{2}\sum_{k=0}^{\infty }{\frac{1}{(k!)^{2}}\frac{(1-\rho_e^2) ^{k}}{(\rho_e)^{2k+1}}} \mathrm{e}^{\left[ -\frac{1}{\rho_e^2}\left(\frac{y_{1}^{\beta}}{\mu _{1}}+\frac{y_{2}^{\beta}}{\mu _{2}}\right)\right]}{\frac{(y_{1}y_{2})^{-1+(k+1)\beta }}{(\mu _{1}\mu _{2})^{k+1}}}, \label{eq_joint}
\end{align}
%%%%%%%%%%%%%%%%%% 
where $\mu_j=E\{V_j\}$ and $j \in (1,2)$ and $\beta$ is their Weibull shape parameter. 
%%%%%%%%%%%%%%%%%%%
\section{Outage probability analysis}
Following section provides analysis of outage probability. 
\vspace{-3mm}
\subsection{Secrecy Outage Probability under Outdated CSI}
The achievable rate for the main and wiretap links is written as $\tilde{{C}}_s\left(i\right)={\mathrm{log}}_{\mathrm{2}}\left(1+{{\tilde{\gamma} }}_{is}\right)$ and ${C}_e\left(i\right)={\mathrm{log}}_{\mathrm{2}}\left(1+{{\gamma}}_{ie}\right)$, respectively.\footnote{We quantify performance penalty in transmission schemes using outdated eavesdropper CSI such that for perfect CSI or schemes like round-robin that do not require this CSI, $\rho=1$ to give ${h}_{ie}=\tilde{h}_{ie}$ in the system model.} The achievable secrecy rate is then defined as the non-negative difference between these rates and is expressed as \cite{zou2016intercept}
%%%%%%%%%%%%%%%%%%
\begin{equation}
\tilde{C}_{sec}\left(i\right)=\max\left\{\,\tilde{C}_s\left(i\right)-{C}_e\left(i\right),0\,\right\},\label{eq_secrecy_capacity_def}
\end{equation}
%%%%%%%%%%%%%%%%%%

A secrecy outage event occurs when $\tilde{C}_{sec}\left(i\right)$ falls below some target secrecy rate $R_s>0$. The secrecy outage probability (SOP) is then expressed as 
%%%%%%%%%%%%%%%%%%
\begin{equation}
\tilde{P}_{out,i}=\mathrm{Pr}\left(\tilde{C}_{sec}\left(i\right)<R_s\right)=1-\tilde{P}_{cov,i},\label{eq_pout_def_2}
\end{equation}
%%%%%%%%%%%%%%%%%%
where $\tilde{P}_{cov,i}\triangleq\mathrm{Pr}\left(\tilde{C}_{sec}\left(i\right)>R_s\right)$ is the coverage probability, which by using (\ref{eq_secrecy_capacity_def}) can be written as
\begin{align}
\tilde{P}_{cov,i}&=\mathrm{Pr} \left({{\tilde{\gamma}}}_{is}>2^{R_s}\left(1+{\gamma}_{ie}\right)-1\right).\label{eq_pcov_def_1}
\end{align}
%%%%%%%%%%%%%%%%%%

Then after some mathematical manipulations (\ref{eq_pcov_def_1}) can be further evaluated as
%%%%%%%%%%%%%%%%%%
\begin{equation}
\tilde{P}_{cov,i}=\int^{\infty }_0{\left[1-F_{{\tilde{\gamma}}_{is}}\left(2^{R_s}\left(1+{\gamma }_{ie}\right)-1\ \right)\right]\ }f_{{\gamma}_{ie}}({\gamma }_{ie})\ d{\gamma }_{ie},\label{eq_pcov_def_4}
\end{equation}
%%%%%%%%%%%%%%%%%%
where $F_{{\tilde{\gamma}}_{is}}\left(.\right)$ denotes the cumulative distribution function (CDF) of $\tilde{\gamma}_{is}$. Now using (\ref{eq_joint}) and \cite[Eq.(8.310),(8.350)]{gradshteyn2014table}, the marginal distribution $f_{\gamma_{ie}}(\gamma_{ie})$ is derived after some algebraic manipulations as
%%%%%%%%%%%%%%%%%%
\begin{align}
f_{\gamma_{ie}}(\gamma_{ie})&=\beta _{e}\sum_{k=0}^{\infty}\sum_{m=0}^{k}{\frac{(\rho_e )^{-m}}{m!}}\times\frac{(1-\rho_e^2 )^{k}}{k!(\rho_e)^{k}}\times\frac{e^{\left[ -\frac{1}{\rho_e}\left(\frac{\Gamma (\alpha _{e})\gamma_{ie}}{\bar{\gamma }_{ie}}\right)^{\beta _{e}}\right] }}{\left(\frac{\bar{\gamma}_{ie}}{\Gamma (\alpha _{e})}\right)^{\beta _{e}(k+1)}}\times\frac{\gamma_{ie}^{\beta _{e}(m+k+1)-1}}{(\frac{\bar{\gamma}_{ie}}{\Gamma (\alpha_{e})})^{\beta_{e}m}}
\end{align}
%%%%%%%%%%%%%%%%%%
where $\alpha_e=\left(1+\frac{1}{\beta_e}\right)$ and $\beta_e$ is the Weibull shape parameter of wiretap link. Furthermore $\frac{\bar{\gamma}_{ie}}{\Gamma(\alpha_e)}$ is the Weibull scale parameter of wiretap link; $\bar{\gamma}_{ie}$ represents the mean value and $\Gamma(.)$ is the well-known Gamma function. 

Since the instantaneous SNR of the sensor nodes is statistically independent, therefore, the CDF of $\tilde{\gamma}_{is}$ can be written as $F_{{\tilde{\gamma}}_{is}}(x)=1-\mathrm{e}^{-\left(\frac{\Gamma \left(\alpha_s\right)x}{\bar{{\tilde{\gamma}}}_{is}}\right)^{\beta_s}}$ \cite{sagias2005correlated}, where $\beta_s$ is the Weibull shape parameter of main link, $\frac{\bar{{\tilde{\gamma}}}_{is}}{\Gamma(\alpha_s)}$ is the Weibull scale parameter of main link and $\bar{{\tilde{\gamma}}}_{is}$ represents the mean value. Substituting $f_{\gamma_{ie}}(\gamma_{ie})$ and $F_{{\tilde{\gamma}}_{is}}(x)$ into (7) and using (5), after some simplifications the outage probability can be expressed as 
%%%%%%%%%%%%%%%%%%%
\begin{align}
\tilde{P}_{out,i}&=1-\sum_{k=0}^{\infty }{\sum_{m=0}^{k}{\frac{\beta _{e}(\rho_e)^{-m}(1-\rho_e ^{2})^{k}}{m!k!(\rho_e )^{k}}}}\int_{0}^{\infty }{e^{\lbrack-\Theta \rbrack }\chi -e^{\lbrack -\Phi \rbrack }\chi }d\gamma_{ie},
\end{align}
%%%%%%%%%%%%%%%%%%%
where $\Theta =\left[ \frac{\Gamma (\alpha _{s})(2^{R_{s}}\gamma_{ie}+\gamma_{th})}{\bar{\tilde{\gamma} }_{is}}\right] ^{\beta _{s}}$, $\Phi =\frac{1}{\rho_e }\left(\frac{\Gamma (\alpha _{e})\gamma_{ie}}{\bar{\gamma}_{ie}}\right)^{\beta _{e}}$, $\chi =\frac{\gamma_{ie}^{\beta_{e}(m+k+1)-1}}{\left(\frac{\bar{\gamma}_{ie}}{\Gamma (\alpha_{e})}\right)^{\beta_{e}(m+k+1)}}$ and $\gamma_{th} = 2^{R_s}-1$. The outage performance is readily evaluated by (9) using any standard computational software; accurate results are achieved by sufficiently considering the first 10 terms of the infinite sum. 
%%%%%%%%%%%%%%%%%%%%
\subsection{Round-Robin Scheduling}
The round-robin scheduling scheme allows $N$ sensor nodes to take turns in transmitting their data to sink. The outage probability in one transmission burst is given by the mean outage probability as \cite{zou2016intercept}
%%%%%%%%%%%%%%%%%%
\begin{equation}
\tilde{P}^{round-robin}_{out}=\frac{1}{N}\sum^N_{i=1}{{\tilde{P}_{out,i}}}.\label{eq_pout_rr_def}
\end{equation}
%%%%%%%%%%%%%%%%%%
\subsection{Best-Node Scheduling Scheme}
In this scheme, the sink selects the sensor node with the highest achievable secrecy rate for transmission in each channel realization. The SOP of said scheme is written as \cite{zou2016intercept}
%%%%%%%%%%%%%%%%%%
\begin{equation}
\tilde{P}^{best}_{out}=\mathrm{Pr}\left[\max_{i\in S}\mathrm{log}_2\left(\frac{1+{\tilde{\gamma}}_{is}}{1+\gamma_{ie}}\right)<R_s\right]=\prod^N_{i=1}{\tilde{P}_{out,i}}.\label{eq_pout_best_def_2}
\end{equation}
%%%%%%%%%%%%%%%%%%
\subsection{Outage Conditioned on Successful Transmission}
The foregoing SOP analysis does not differentiate between reliable and secure communications - outage is declared either when the sink cannot successfully decode the information or when some information is leaked to the eavesdropper. We now characterize SOP under Weibull fading for the latter case, i.e., analyze the probability of a decoded message failing to achieve perfect secrecy \cite{zhou2011rethinking}. Consider that the encoder at $i$-th node employs an $\left(R_{it},R_s\right)$ wiretap code of block length $L$, where $R_{it}$ is the rate of transmitted codewords and $R_s$ is the rate of embedded secrecy information bits \cite{zhou2013physical}. For each message index $p$, a codeword $x^L(p,q)$ is transmitted where $p=1,2,...2^{LR_s}$ and for each $p$ the index $q$ is drawn according to a uniform distribution from $q=1,2,...2^{LR_{it}}$ \cite{zhou2013physical}. The difference $(R_{it}-R_s)$, known as rate redundancy, is a measure of the information rate sacrificed to introduce randomness for providing secrecy against the eavesdropper. For a message transmitted by the $i$-th node, the sink can decode correctly as long as $R_{it}\leq \tilde{C}_s(i)$ whereas perfect secrecy is compromised if $\left(R_{it}-R_s\right)<{C}_e\left(i\right)$. Then the probability of a decoded message at sink failing to achieve perfect secrecy under non adaptive encoding, is written as \cite[Eq.(8)]{zhou2011rethinking}
%%%%%%%%%%%%%%%%%%
\begin{align}
\tilde{P}_{sop,i}&=\mathrm{Pr}\left\{\left.{C}_e\left(i\right)>R_{it}-R_s\,\right|\textrm{successful decoding}\right\}\nonumber\\
&={\mathrm{Pr}\left\{{C}_e\left(i\right)>R_{it}-R_s\right\}} \nonumber \\
&=F_{\gamma_{ie}}(2^{R_{it}-Rs}-1),\label{eq_psop_i_def_1}
\end{align}
%%%%%%%%%%%%%%%%%%
where the cdf of $\gamma_{ie}$ is obtained by using (8) and \cite[Eq.(8.350),(8.352)]{gradshteyn2014table} as
%%%%%%%%%%%%%%%%%%
\begin{align}
F_{\gamma_{ie}}(x)&=\rho_e \sum_{k=0}^{\infty}\frac{(1-\rho_e ^{2})^{k}}{k!}\gamma\left(k+1,\frac{1}{\rho_e }\left(\frac{\Gamma(\alpha _{e})x}{\bar{\gamma}_{ie}}\right)^{\beta_{e}}\right),
\end{align}
%%%%%%%%%%%%%%%%%%
where $\gamma(.,.)$ is the incomplete Gamma function. It may be noted that the outage probability derived in (9) does not differentiate between reliable and secure communications such that an outage event occurs either when sink cannot decode the information or when information is leaked to the eavesdropper. In contrast, the derivation of (12) considers information leakage to the eavesdropper conditioned on the sink reliably decoding the message. Therefore, (12) provides a more explicit measure of the secrecy outage performance.
%%%%%%%%%%%%%%%%%%%%%%%%%%%%
\section{Numerical results}
This section provides numerical examples for the derived analytical results, which are also validated by simulations. 
%%%%%%%%%%%%%%%%%%
%\vspace{-3mm}
\begin{figure}[htb]
	\centering
		\includegraphics[scale=0.31]{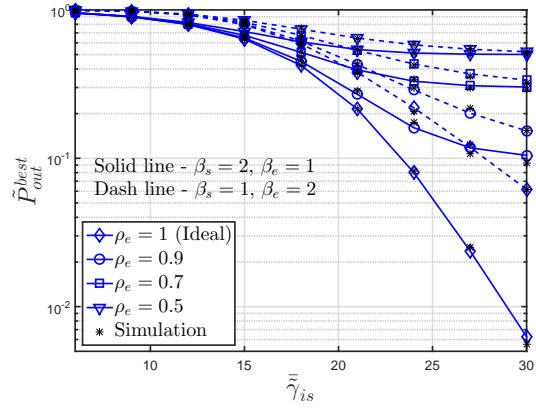}
	\caption{Outage probability against $\bar{{\tilde{\gamma}}}_{is}$. Other parameters are $N=1$, $R_s=1$ bit/s/Hz and $\bar{\gamma}_{ie}$ = 15 dB.}
	\label{fig:fig11}
\end{figure}
%\vspace{-5mm}
%%%%%%%%%%%%%%%%%%%%
\begin{figure}[htb]
	\centering
		\includegraphics[scale=0.31]{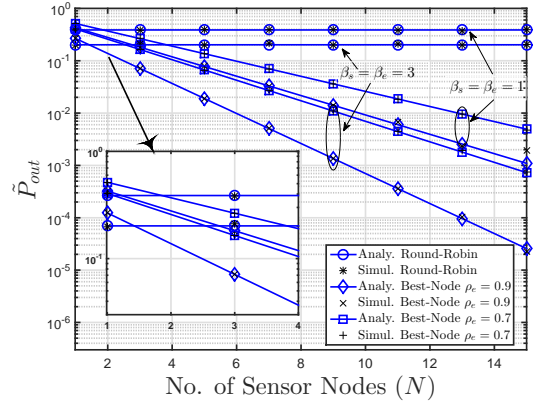}
	\caption{$\tilde{P}_{out}$ for both scheduling schemes versus network size, $N$. Other parameters are $R_s=1$ bit/s/Hz, $\bar{{\tilde{\gamma}}}_{is}=20$ dB and $\bar{\gamma}_{ie}=15$ dB.}
	\label{fig:fig22}
\end{figure}
%%%%%%%%%%%%%%%%%%

Fig. \ref{fig:fig11} shows how the outage probability is affected by the non-identically distributed Weibull fading between the main and wiretap links; $\tilde{P}^{best}_{out}$ is plotted as a function of increasing values of main link SNR for different values of $\beta_s$ and $\beta_e$. It can be seen from the figure that for a given SNR if the main channel’s shape parameter $\beta_s$ is smaller than  $\beta_e$ then outage probability increases, which corresponds to a reduced secrecy performance. This corroborates with more severe fading on the main link relative to the wiretap link when $\beta_s<\beta_e$. From the same figure, one may also observe that an increase in $\rho_e$ decreases the outage probability.

Fig. \ref{fig:fig22} plots outage probability for the best-node selection and the round-robin scheduling schemes as a function of network size, $N$. It can be observed from the figure that for the best-node scheduling scheme $\tilde{P}_{out}$ reduces with increasing $N$ due to it being a product of $N$ terms, each smaller than unity, as shown in (\ref{eq_pout_best_def_2}). Furthermore, for diminishing $\rho_e$ values the best-node scheme's outage probability versus $N$ graph has a progressively smaller slope, i.e., the advantage gained by increasing network size is significantly reduced as the eavesdropper CSI becomes more outdated. Specifically, for $N=5$ the outage probability of best node scheduling increases from 0.01 to 0.06 when $\rho_e$ decreases from 0.9 to 0.7 for $\beta_s=\beta_e=3$. Moreover, Fig. \ref{fig:fig22} also shows that best-node scheduling is more sensitive to $\rho_e$ when $\beta_s=\beta_e>1$.  
%%%%%%%%%%%%%%%%%%%%%%
%\vspace{-1.5mm}
\begin{figure}[htb]
	\centering
		\includegraphics[scale=0.31]{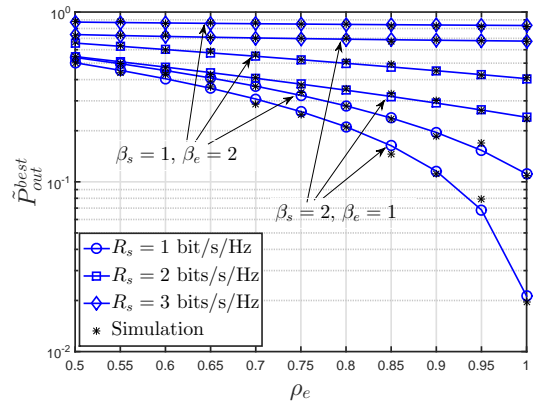}
	\caption{$\tilde{P}_{out}^{best}$ versus $\rho_e$. Other parameters are $N=1$, $\bar{{\tilde{\gamma}}}_{is}=20$ dB and $\bar{\gamma}_{ie}=0$ dB.}
	\label{fig:fig33}
\end{figure}
%\vspace{-5mm}
%%%%%%%%%%%%%%%%%%
\begin{figure}[htb]
	\centering
		\includegraphics[scale=0.31]{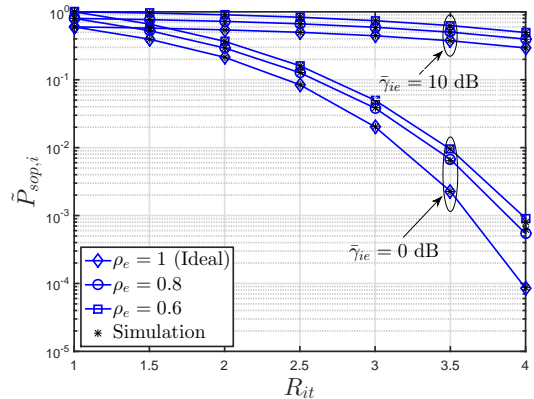}
	\caption{$\tilde{P}_{sop,i}$ against $R_{it}$. Other parameters are $R_s=1$ bit/s/Hz.}
	\label{fig:fig44}
\end{figure}
%%%%%%%%%%%%%%%%%%

Fig. \ref{fig:fig33} plots outage probability for the best-node scheme against increasing $\rho_e$ values. The figure shows that the outage probability generally decreases as $\rho_e \rightarrow 1$, but it decreases more rapidly when $\beta_s>\beta_e$; for $\beta_s<\beta_e$, i.e., more severe fading on the main link, a reduction in $\rho_e$ causes a gentler decline in outage probability. This shows that the wiretap link correlation has more effect on outage performance when the main link fading severity is more than that of the wiretap link. Moreover, the outage probability also increases with increasing $R_s$ but the impact of varying $\rho_e$ is less prominent at higher values of $R_s$.

In Fig. \ref{fig:fig44} we have plotted the conditional SOP against increasing values of $R_{it}$, for $R_s=1$, to elaborate the dependence of secrecy outage on codeword transmission rate. It can be observed from the figure that by increasing $R_{it}$, the conditional SOP is reduced as there is more randomness in the codeword, indicated by the difference $(R_{it}-R_s)$, to provide secrecy against the eavesdropper. Also, for large $R_{it}$ the impact of $\rho_e$ on the conditional SOP becomes more prominent due to the enhanced reliability at the cost of security. Moreover, we observe that as the SNR of the eavesdropper link becomes larger, the impact of $\rho_e$ reduces, hence compromising the security of the wireless link.
%\vspace{-0.5mm}
\section{Conclusion}
We have analyzed physical layer security for a WSN under Weibull fading and imperfect estimate of the eavesdropper channel. Our results demonstrate that for non-identically distributed Weibull fading on the main and wiretap links, it is desirable to have $\beta_s>\beta_e$ to achieve a smaller value of the outage probability for a given network size $N$. Furthermore, we have quantified the effect of outdated eavesdropper channel state on the best-node scheme and compared it with the round-robin scheme. For round-robin scheduling, the outage probability is practically un-changed for increasing $N$, while the best-node scheme's outage probability decreases for increasing $N$ at a much faster rate. Although, the best-node scheme's outage probability increases with more outdated eavesdropper CSI, but for network size larger than 5 it outperforms the round-robin scheme. The SOP conditioned on successful decoding has been analyzed separately and it is shown that when more randomness can be included in the code then the impact of outdated eavesdropper CSI on secrecy outage probability also diminishes. Our results are useful for analyzing secrecy outage for WSNs that experience Weibull fading.
\vspace{-1.5mm}
% if have a single appendix:
%\appendix[Proof of the Zonklar Equations]
% or
%\appendix  % for no appendix heading
% do not use \section anymore after \appendix, only \section*
% is possibly needed

% use appendices with more than one appendix
% then use \section to start each appendix
% you must declare a \section before using any
% \subsection or using \label (\appendices by itself
% starts a section numbered zero.)

% use section* for acknowledgment
%
%
%\section*{Acknowledgment}
%This work is supported by the EU-funded project ATOM-690750, approved under call H2020-MSCA-RISE-2015.
%
%
% Can use something like this to put references on a page
% by themselves when using endfloat and the captionsoff option.
\ifCLASSOPTIONcaptionsoff
  \newpage
\fi
% trigger a \newpage just before the given reference
% number - used to balance the columns on the last page
% adjust value as needed - may need to be readjusted if
% the document is modified later
%\IEEEtriggeratref{8}
% The "triggered" command can be changed if desired:
%\IEEEtriggercmd{\enlargethispage{-5in}}

% references section

% can use a bibliography generated by BibTeX as a .bbl file
% BibTeX documentation can be easily obtained at:
% http://mirror.ctan.org/biblio/bibtex/contrib/doc/
% The IEEEtran BibTeX style support page is at:
% http://www.michaelshell.org/tex/ieeetran/bibtex/
%\bibliographystyle{IEEEtran}
% argument is your BibTeX string definitions and bibliography database(s)
%\bibliography{IEEEabrv,../bib/paper}
%
% <OR> manually copy in the resultant .bbl file
% set second argument of \begin to the number of references
% (used to reserve space for the reference number labels box)
%%%%%%%%%%%%%%%%%%%%%%%%%%%%%%
\bibliographystyle{IEEEtran}% This is IEEEtran.bst file
%\bibliography{./IEEEfull,./Alvi_Nakagami_Fading}% Use this reference database
%\bibliography{IEEEabrv,Furqan_Weibull}
\bibliography{IEEEabrv,CL2017_0561}
\end{document}